# Orbitally-driven Behavior: Mott Transition, Quantum Oscillations and Colossal Magnetoresistance in Bilayered $Ca_3Ru_2O_7$

G. Cao[1], X.N. Lin[1], L. Balicas[2], S. Chikara[1], J.E. Crow[2,3] and P. Schlottmann[2,3]

[1]Department of Physics and Astronomy, University of Kentucky, Lexington, KY 40506

[2] National High Magnetic Field Laboratory, Tallahassee, FL 32310

[3]Department of Physics, Florida State University, Tallahassee, FL 32306

We report recent transport and thermodynamic experiments over a wide range of temperatures for the Mott-like system $Ca_3Ru_2O_7$ at high magnetic fields, B (≤ 30 T). This work reveals a rich and highly anisotropic phase diagram, where applying B along the *a-, b-,* and *c*-axis leads to vastly different behavior. A fully spin-polarized state via a first order metamagnetic transition is obtained for B ≥ 6 T and B||*a*, and colossal magnetoresistance is seen for B||*b*, and quantum oscillations in the resistivity are observed for B||*c*, respectively. The interplay of the lattice, orbital and spin degrees of freedom are believed to give rise to this strongly anisotropic behavior.

## I. Introduction

The ruthenates belong to a class of new materials of highly correlated electrons that is rich in novel physical properties. It has become increasingly clear that in the ruthenates the orbital degrees of freedom play a fundamental role. The phenomena are driven by the coupling of the orbits to the spin (spin-orbit interaction) and to the lattice (Jahn-Teller effect). The central feature of the 4d-electron based materials is their extended orbitals, which lead to comparable and hence competing energies for the crystalline fields, Hund's rule interactions, spin-orbit coupling, p-d hybridization and electron-lattice coupling. The Ru ions are surrounded by six O-atoms forming $RuO_6$ octahedra. Crystallographic distortions and the relative orientation of the octahedra and their tilting are believed to determine the exotic properties of $Ca_3Ru_2O_7$ and other ruthenates.

The bilayered $Ca_3Ru_2O_7$ features a Mott-like transition, a metamagnetic transition [1, 2], colossal magneto-resistance and quantum oscillations in the resistivity [3-5]. Due to the strong crystalline field anisotropies, the coupling of the magnetic field to the system crucially depends on the orientation of the field relative to the crystal axis. This is evidence for the importance of the orbital degrees of freedom in this system and other ruthenates. The Shubnikov-de Haas oscillations for fields along the *c*-axis confirm the existence of a well-defined Fermi surface with closed orbits in the *ab*-plane [3]. The poor metallic behavior of the $Ca_3Ru_2O_7$ system observed in the transport properties is then a consequence of the low carrier density.

For temperatures below 56 K and small magnetic fields the magnetic properties of $Ca_3Ru_2O_7$ are consistent with antiferromagnetic long range order. The ordered magnetic moments lie in the *ab*-plane, but the details of the relative spin-orientations is not yet established. When the temperature is lowered to 48 K a discontinuous transition leads to a new state with a partially gapped Fermi surface. The first order transition manifests itself in the magnetization, the lattice parameter along the *c*-axis, the resistivity and the Raman scattering spectrum. We speculate that this new phase has both, antiferromagnetic and orbital, long-range order and that the charge gap of 0.1 eV observed by Raman spectroscopy [6, 7] is due to the orbital order rather than of the Mott-type.

As a function of field, a metamagnetic transition is observed when a field of 6 T is applied along the *a*-axis. For fields larger than 6 T the spins are ferromagnetically aligned. The transition is also accompanied by a colossal magnetoresistive effect. The ferromagnetic alignment of the spin gives rise to a reduction in the resistivity by one order of magnitude. The physics is very different if the magnetic field is applied along the *b*- or *c*-axis.

In this paper, we present the experimental evidence for a yet incomplete phase diagram. In section II we show magnetization and resistivity measurements as a function of temperature and a wide range of magnetic fields ($0 \leq B \leq 30$ T) applied along the *a*-, *b*- and *c*-axis. The results are discussed in section III in an attempt to provide a coherent picture for the bilayer $Ca_3Ru_2O_7$. It is now clear that the lattice, orbital and spin degrees of freedom are intimately coupled to each other. A brief summary is given in section IV.

## II. Results

The bilayered $Ca_3Ru_2O_7$ is the second member of the Ruddlesden-Popper series with lattice parameters of *a*=5.3720(6) Å, *b*=5.5305(6) Å, and *c*=19.572(2) Å [2]. The crystal structure is severely distorted by tilting of the $RuO_6$ octahedra. The tilt projects primarily onto the *ac*-plane (153.22°), while it only slightly affects the *bc*-plane (172.0°) [2]. These are crucial bond angles determining the crystalline field splitting of the 4d-orbitals and the overlap matrix elements between orbitals within the basal plane. They directly impact the band structure and are the origin for the anisotropic properties of the compound. $Ca_3Ru_2O_7$ displays antiferromagnetic (AFM) long-range order below $T_N$=56 K while remaining metallic [1], and then at $T_{MI}$=48 K it undergoes a partial Mott-like transition [1-10]. This partial Mott transition is associated with a dramatic reduction in the conductivity for $T < T_{MI}$ and the formation of a charge gap of 0.1 eV [6, 7] in one of the bands, while the system still remains a poor metal.

Shown in Fig. 1(a) is the temperature dependence of the magnetization, M(T), for a field of B=0.5 T applied along the *a*-, *b*- and *c*-axis (left scale) together with the temperature dependence of the lattice parameter of the *c*-axis (right scale). M(T) for the three orientations at B$\geq$6.5 T is presented in Fig. 1(b). The magnetization strongly depends on the orientation of the magnetic field. There are two major features that are critically linked to later discussions.

*First,* the low field M(T) for the *a*-axis, the magnetic easy axis, shows two phase transitions, $T_N$=56 K and $T_{MI}$=48 K. In contrast, M(T) for the *b*-axis exhibits no discernable anomaly corresponding to $T_{MI}$ but a sharp peak at $T_N$ as seen in Fig.1(a). For the c-axis, the two transitions are observed but considerably weakened. In addition, the precipitous decrease of M for the *a*-axis at $T_{MI}$ when T is lowered is accompanied by a simultaneous sudden reduction in the *c*-axis lattice parameter. Hence, there is substantial magnetoelastic coupling that leads to Jahn-Teller distortions of the $RuO_6$ octahedra and their relative orientations [2, 3].

*Secondly,* as B applied parallel to the *a*-axis increases, $T_{MI}$ shifts slightly downward, whereas $T_N$ remains initially essentially unchanged and eventually becomes rounded off at higher fields. For B≥6 T, the magnetic state is driven to a *spin-polarized or ferromagnetic* (FM) state. In contrast, when B is parallel to the *b*-axis, $T_N$ decreases with increasing B approximately at a rate of 2K/T. Remarkably, the magnetic ground state for B||*b*-axis *remains antiferromagnetic*, entirely different from that for B||*a*-axis (see Fig. 1(b)). Unlike the *a*- and *b*-axis magnetization, the *c*-axis magnetization remains essentially unchanged. This suggests that intraplane spin couplings are particularly strong.

The anisotropy of the magnetic state is further illustrated in Fig.2 showing the isothermal magnetization at T=5 K. It displays the first-order metamagnetic transition at B=6 T leading to the spin-polarized or ferromagnetic state with a saturation moment $M_s$ =1.73 $\mu_B$/Ru when B is applied along the *a*-axis. Hence, more than 85% of the hypothetical saturation magnetization of 2 $\mu_B$/Ru expected for an S=1 system is achieved. The behavior is completely different if the field is applied along the *b* or *c*-axis, in part due to a strong anisotropy field of 22.4 T [10].

Fig.3(a) and (b) show the resistivity for the current along the *c*- and *a*-axis, $\rho_c$ and $\rho_a$, respectively, as a function of B for B||*a*-, *b*- and *c*-axis at T=0.6 K. Both, $\rho_c$ and $\rho_a$, drastically depend on the orientation of the field. This anisotropy and the coupling of the spin, orbital and lattice degrees of freedom are a central finding of this work. For B||*a*-axis *(magnetic easy-axis)* both $\rho_a$ and $\rho_c$ show an abrupt drop by an order magnitude at 6 T, which corresponds to the first-order metamagnetic transition leading to the spin-polarized state with more than 85% polarized spins as shown in Fig.2. Further increases of B up to 30 T only result in a weak linear dependence of the resistivity with B. In contrast, for B||*b*-axis *(magnetic hard-axis),* where the magnetic state remains antiferromagnetic, $\rho_c$ and $\rho_a$ rapidly decrease by as much as two and three orders of magnitude, respectively, at a critical field of $B_c$=15 T ($B_c$ decreases with increasing temperature). For *B||c-axis,* both, $\rho_c$ and $\rho_a$, show Shubnikov-de Haas oscillations corresponding to very small Fermi surface cross sections.

It is clear that the transport properties strongly depend on the magnitude and orientation of the magnetic field. Such anisotropic coupling of the field to the system is only conceivable via orbital degrees of freedom coupled via the spin-orbit interaction to the spin of the 4d-electrons. It also requires that the degeneracy of the $t_{2g}$ orbitals be lifted by deformations of the $RuO_6$ octahedra. In particular the tilting angle of the octahedral provides a sensitive coupling to the lattice. Hence, spin, orbital and lattice degrees of freedom are intimately coupled with each other. A reduction of the resistivity with the magnetization is known from colossal magnetoresistive (CMR) materials such as the manganites [11,12] and this mechanism could explain the metal-insulator transition for B||a. But the physics in $Ca_3Ru_2O_7$ is more complex from that driving other magnetoresistive materials, where a spin-polarized state is essential for CMR [12]. This is evidenced in Fig. 3 with the strong decrease of the resistivity for B||b, while the magnetization is small and proportional to the field (see Fig. 2). The question is then, *what is the origin of the abrupt reduction of the resistivity by as much as three orders of magnitude when B||b-axis*? Below we address this issue with more detailed results.

*A. Colossal magnetoresistance* -- Shown in Fig.4 is the temperature dependence of $\rho_c$ with B||*a*-axis (Fig. 4(a)) and B||*b*-axis (Fig. 4(b)) for a few representative values of B (Note that the same log scale is used in both panels to facilitate comparison). $\rho_a$

displays the same temperature dependence for B||*a*- and *b*-axis, and is therefore not shown here. For B||*a*-axis, at low temperatures $\rho_c$ increases slightly with increasing B when B<6 T, and decreases abruptly by about an order of magnitude when B≥6 T, at which the first order metamagnetic transition leads to the spin-polarized state as seen in Figs. 2 and 3. A further increase in B results in a slightly higher resistivity at low temperatures. $\rho_c$ corresponds to the coherent motion of electrons between Ru-O planes separated by insulating (I) Ca-O planes. A reduction of $\rho_c$ in high fields is then attributed to a field-induced enhancement of the tunneling between the planes due to the spin-polarization. This situation is similar to an array of FM/I/FM junctions or spin-filters where the probability of tunneling and thus the electronic conductivity depends on the angle between the spin magnetization of adjacent ferromagnets. For $Ca_3Ru_2O_7$ the spin-polarized state still does not generate~~s~~ a fully metallic state in spite of the pronounced reduction of $\rho_c$. In fact, the $\rho_c$ at B=28 T is still that of a bad metal or low carrier density system (see the inset). In sharp contrast, when B||*b*-axis, the magnetic hard axis, the temperature of the sharp drop in $\rho_c$ decreases approximately at a rate of 2K/T, and disappears for B>24 T, as seen in Fig. 4(b). Again, these results suggest a strong magneto-orbital coupling.

*B. Rotation of easy axis* – As shown in Fig. 5, in the vicinity of $T_{MI}$ (42 $\leq$ T $\leq$48 K), the magnetic easy-axis starts to rotate away from the *a*-axis and becomes parallel to the *b*-axis close to $T_{MI}$ with the saturation moment $M_s$ being only 0.8 $\mu_B$/Ru. Here the angle θ is defined as the angle between B and the *a*-axis and the applied field is of 6.2 T. This rotation of the easy axis is driven by the change of the c-axis lattice parameter with temperature shown in Fig.1. Between $T_{MI}$ and $T_N$ (the shaded area in Fig.1) the ferromagnetism does no longer favor the *a*-axis, but the *b*-axis. Hence, the magnetic states below $T_{MI}$ and between $T_{MI}$ and $T_N$ evolve out of a different antiferromagnetically ordered state.

*C. Quantum oscillations (B||c-axis)*—When B is applied along the *c*-axis, the quantum oscillations are observed in both $\rho_a$ and $\rho_c$ for 20mK<T<6.5 K. Illustrated in Fig. 6(a) is $\rho_c$ as a function of B ($\rho_a$ is not shown). The amplitude of the quantum oscillations as a function of inverse field $B^{-1}$ is presented in Fig. 6(b) and 6(c) for $\rho_c$ and $\rho_a$, respectively. Quantum oscillations are a trademark of a Fermi liquid with closed orbits and long mean free path. The Shubnikov-de Haas oscillations are found for B||*c*-axis, but no quantum oscillations are discerned for B||*a*- or *b*-axis. Hence, they must be associated with the motion of the electrons in the *a*-*b* planes.

The analyses of the oscillations for $\rho_c$ reveal an extremely low frequency $f_1$ = 28 T, which, based on crystallographic data [2] and the Onsager relation $F_0 = A(h/4\pi^2 e)$ (*e* is the electron charge, *h* Planck constant), corresponds to an area of only 0.2% of the first Brillouin zone. Shown in Fig.6b as the thin lines is the beating between two frequencies ($f_1$=28 T and $f_2$=10 T). The same analyses for $\rho_a$ yields a slightly higher frequency of 33 T, shown in Fig.6(c). The similar frequencies estimated from both $\rho_a$ and $\rho_c$ imply that the same orbitals may be responsible for the quantum oscillations. The cyclotronic effective mass is estimated to be $\mu_c$=0.85 ± 0.05. It is markedly smaller than the enhanced thermodynamic effective mass (~3) estimated from the electronic contribution, γ, to the specific heat [5]. There are two possible sources for this discrepancy: (1) The cyclotronic effective mass is measured in a relatively large magnetic field that quenches correlations, while the specific heat is a zero-field measurement, and (2) $\mu_c$ only refers to one closed orbit, while the thermodynamic effective mass measures an average over the entire Fermi surface.

In Fig. 7 we present $\rho_c$ as a function of B at low temperature for the field pointing slightly away from the *c*-axis (within the *ac*-plane). As discussed before a field component along the *a*-axis induces the first-order metamagnetic transition. If $\Theta$ is now the angle the field forms with the *c*-axis, then for $\Theta=90^o$ the transition occurs at 6 T, but for smaller $\Theta$ the transition occurs at a larger field. From the previous discussion we expect to see Shubnikov-de Haas oscillations for fields less the field of the transition. However, the oscillations persist for fields above the critical field, but with a larger frequency of 47 T for $\Theta=27^o$ as shown in Fig.7(b). In contrast, surprisingly, no oscillations are discerned for B in the *bc*-plane.

Note that the oscillation signature decreases with increasing angle and, as shown in Fig.7, the quantum oscillations disappear when $\Theta>45^o$, i.e., for the quantum oscillations to occur, the angle between B and the *c*-axis should not be larger than $45^o$. This suggests that for small $\Theta$ the cross section of the closed orbit is slightly enhanced due to the tilting of the direction of the field, but the Fermi surface is otherwise not drastically modified. However, with increasing $\Theta$ the effect of the field on the Fermi surface is more dramatic and closed orbits disappear; they are possibly replaced by open ones, which do not contribute to quantum oscillations.

**III. Discussion**

The unusual and novel behavior observed in the ruthenate $Ca_3Ru_2O_7$ is predominantly associated with the role of the orbital degrees of freedom and their coupling to the electron spin and the lattice. The orbital degeneracy of the 4d levels is lifted by the crystalline field splitting arising in first place by the $RuO_6$ octahedra. In first approximation the two $e_g$ levels are split off from the three $t_{2g}$ orbitals. The Hund's rule energy maximizing the total spin at each Ru site is not large enough to overcome the $e_g$-$t_{2g}$ splitting, so that the $e_g$ levels are not populated and can be disregarded for all practical purposes. Hence one $t_{2g}$ orbital is doubly occupied, while the other two host a single electron each. The octahedra are deformed (all lattice parameters are different) and consequently the three $t_{2g}$ levels are expected to have different energies. These splittings are believed to be larger than the thermal energy $k_BT$ and the Zeeman effect.

The octahedra are corner-shared and tilted, with a larger tilting angle in the *ac*-plane than in the *bc*-plane. The tilting plays a fundamental role for the hybridization matrix elements between states corresponding to neighboring octahedra. Small changes in the tilting can give rise to qualitative changes in the properties. In view of the strong crystalline field and tilting angle asymmetries, the coupling of the magnetic field to the system depends on the orientation of the field. In other words, the matrix elements for the orbital Zeeman effect depend on the direction of the field. Consequently, different properties can be expected if the field is applied along the *a*-, *b*- or *c*-axis.

As shown in Fig.1(a), for a small magnetic field, the system is paramagnetic at high temperatures and undergoes a transition to an antiferromagnetic state at $T_N$ = 57 K. The ordered magnetic moment lies in the *a*-*b* plane and appears to be oriented along the *a*-axis. A magnetic field of about 6 T along the *a*-axis flops the spins of one of the antiferromagnetic sublattices leading to a ferromagnetic state. No such reorientation effect is observed if the field is applied along the *b*- or *c*-axis.

Also shown in Fig. 1(a) is an abrupt decrease in the *c*-axis lattice parameter at $T_{MI}$ as the temperature is lowered. This change in the lattice parameter is accompanied by a decrease in the magnetization for the field applied along the *a* and *c* directions (but not along *b*) and an increase of the resistivity $\rho_c$. The transition temperature $T_{MI}$ decreases with increasing field if the field is applied along the *b*-axis. A Raman-scattering study of $Ca_3Ru_2O_7$ revealed that the transition at $T_{MI}$ is associated with the opening of a charge gap, $\Delta_c \sim 0.1$ eV, and the concomitant softening and broadening of an out of phase O phonon mode. This is indicative of a rearrangement in the $t_{2g}$ orbitals, possibly either with a change in the Jahn-Teller distortions, a buckling of Ru-O-Ru bonds or the onset of long-range orbital ordering. The interpretation in terms of orbital order is particularly appealing because of the decrease of $T_{MI}$ with field. Loosely speaking, due to the antisymmetry of the wave-function, spin ferromagnetism tends to be accompanied by antiferro-orbital order. Orbital ordering in single layered ruthenates has been predicted theoretically [13,14] and was suggested from X-ray data [15].

The observed metal-insulator transition is only a partial one, because the compound remains a poor metal below $T_{MI}$. Poor metallic behavior can arise either from a low carrier density or a short mean free path. The observation of Shubnikov-de Haas oscillations at low temperatures is a clear indication of a well-established Fermi surface with closed orbits and a sufficiently long mean-free path of the carriers. The orbits correspond to very small cross-sections of the Fermi surface (less than 1% of the Brillouin zone cross-section). Hence, they must be associated with small electron or hole pockets and the system should be characterized as a low carrier-density compound.

The first order antiferromagnetic to ferromagnetic metamagnetic transition as function of magnetic field, when a field of 6 T is applied along the *a*-axis, is also clearly seen as a drop in the resistivity. This transition resembles that of systems displaying colossal magnetoresistance. In the antiferromagnetic state the transport along the *c*-axis is reduced due to the fact that the spins of the layers are not aligned at B=0. The spin-filter effect then prevents the carriers to tunnel through the layers. In a sufficiently strong magnetic field (along the *a*-direction) the spins are reoriented ferromagnetically and coherent motion of the electrons can again take place.

The electronic band structure and magnetism of $Sr_3Ru_2O_7$ has been investigated using the local density approximation (LDA) and the linearized augmented plane wave method by Singh and Mazin [16]. The Sr-based compound is expected to have some common aspects with the Ca-bilayer system. The calculation was carried out in the idealized tetragonal symmetry and in the experimental orthorhombic structure. The $d_{xy}$-orbitals give rise to a cylindrical electron-like sheet centered about the Γ and Z points with weak dispersion along the c-direction, because of the rather small interplanar coupling. The $d_{xz}$ and $d_{yz}$ orbitals, on the other hand, provide flat sheet-like sections with strong nesting. Nesting could be responsible for the antiferromagnetic long-range order or antiferro-orbital order. The Fermi surface within the LDA calculations is found to be very sensitive to small structural changes and shifts in the Fermi energy.

For the single-layer ruthenate $Ca_{2-x}Sr_xRuO_4$ an orbital-selective Mott-insulator transition has been predicted [14]. This theoretical study (based on band structure calculations using the LDA, LDA+U and the Dynamical Mean-Field Approximation methods) explains the insulating and metallic phases with a coexistence of localized and itinerant orbitals, respectively. The metal-insulator transition observed as a function of x is attributed to a crossover of crystalline field levels and the concomitant localization/delocalization of 4d-orbitals. The orbital-selective Mott-transition mechanism has been questioned by a very detailed model calculation for a two-band Hubbard system [17]. A different explanation for the transitions in the single layer system [18] is based on a cooperative occupation of the $d_{xy}$ orbitals and their ferro-orbital ordering. The charge gap observed in bilayer $Ca_3Ru_2O_7$ [6, 7] is not believed to arise from a Mott metal-insulator transition. The possible mechanisms for its origin are a crystalline field level rearrangement due to coupling to phonons (Jahn-Teller distortion), a buckling of Ru-O-Ru bonds or antiferro-orbital order. The latter is the most likely scenario. Further investigations with microscopic probes are necessary to gain deeper insight into this issue.

## IV. Summary

As summarized in Fig. 8, $Ca_3Ru_2O_7$ displays a rich and complex phase diagram. The distortions and tilting of the $RuO_6$ octahedra give rise to highly anisotropic magnetic and orbital properties. We have presented evidence that in this compound the spin, orbital and lattice degrees of freedom are intimately coupled. As a consequence of the anisotropies in the lattice structure, applying the magnetic field B along the *a-, b-,* and *c*-axis leads to very different properties.

Two transitions are observed in the magnetization, resistivity and the lattice parameter of the *c*-axis when the applied field is small. The high temperature phase is paramagnetic and without orbital order. Below $T_N$ the system orders antiferromagnetically, where the detailed orientation of the ordered moments is not yet known. Below the second transition at $T_{MI}$, a charge gap opens, which could be associated with the onset of antiferro-orbital order. A magnetic field reduces the two transition temperatures.

If a field of 6 T is applied along the *a*-axis the system undergoes a first order metamagnetic transition into a ferromagnetic state. Here the antiferromagnetic order is broken up by the field. It is not clear if the orbital order persists. With increasing T there is a gradual crossover into the paramagnetic/orbitally disordered state. If, on the other hand, the magnetic field is applied along the *b*-axis, an abrupt transition into a metallic state occurs at a much larger field. The transition could be associated with the breakdown of the long range orbital order. Finally, when the field is oriented along the *c*-axis, Shubnikov-de Haas oscillations are found at low temperatures. These oscillations correspond to closed orbits in the *ab*-plane with small cross-section. The large resistivity of the system is attributed to a low carrier-density.

**Acknowledgements**

G.C. is grateful to Dr. Zhong Fang for very helpful discussions. The work done at University of Kentucky was supported by NSF grant No. DMR-0240813. This work was supported by the National High Magnetic Field Laboratory through NSF Cooperative Agreement DMR-0084173 and the State of Florida. P.S. acknowledges the support by NSF (grant No. DMR01-05431) and DOE (grant No. DE-FG02-98ER45707).

**References:**

1. G. Cao, S. McCall, J. E. Crow and R. P. Guertin, Phys. Rev. Lett. **78**, 1751 (1997).
2. G. Cao, K. Abbound, S. McCall, J.E. Crow and R.P.Guertin, Phys. Rev. B **62**, 998 (2000).
3. G. Cao, L. Balicas, Y. Xin, E. Dagotto, J. E. Crow, C.S. Nelson and D.F. Agterberg, Phys. Rev. B **67** 060406 (R) (2003).
4. G. Cao, L. Balicas, Y. Xin, J. E. Crow, and C.S. Nelson, Phys. Rev. B **67** 184405 (2003).
5. G. Cao, L. Balicas, X.N. Lin, S. Chikara, V. Durairaj, E. Elhami, J.W. Brill, R.C. Rai and J. E. Crow, Phys. Rev. B **69**, 014404 (2004)
6. H.L. Liu, S. Yoon, S.L. Cooper, G. Cao and J.E. Crow, Phys. Rev. B **60**, R6980, (1999).
7. C.S. Snow, S.L. Cooper, G. Cao, J.E. Crow, S. Nakatsuji, Y. Maeno, Phys. Rev. Lett. **89**, 226401 (2002).
8. A.V. Puchkov, M.C. Schabel, D.N Basov, T. Startseva, G. Cao, T. Timusk, and Z.-X. Shen, Phys. Rev. Lett. **81**, 2747 (1998).
9. R.P. Guertin, J. Bolivar, G. Cao, S. McCall, and J.E. Crow, Solid State Comm. **107**, 263 (1998).
10. S. McCall, G. Cao, and J.E. Crow, Phys. Rev. B **67,** 094427 (2003).
11. For example, E.Y. Tsymbal and D.G. Pettifor, p. 113, *Solid State Physics* v. 56, ed. Henry Ehrenreich and Frans Spaepen (Academic Press, New York, 2001)
12. For example, Yoshinori Tokura, *Colossal Magnetoresistive Oxides* (Gordon and Beach Science Publishers, Australia, 2000).

13. T. Hotta and E. Dagotto, Phys. Rev Lett. **88**, 017201 (2002).

14. V.I. Anisimov, I.A. Nekrasov, D.E. Kondakov, T.M. Rice and M. Sigrist, Eur. Phys. J. B **25**, 191 (2002).

15. T. Mizokawa *et al.*, Phys. Rev. Lett. **87**, 077202 (2001).

16. D.J. Singh and I.I. Mazin, Phys. Rev. B **63**, 165101 (2001).

17. A. Liebsch, Phys. Rev. Lett. **91**, 226401 (2003).

18. Z. Fang, N. Nagaosa, and K. Terakura, Phys. Rev. B **69**, 045116 (2004).

**Captions:**

**Fig.1.** Temperature dependence of magnetization, M, for the field along the *a*-, *b*- and *c*-axis for (a) B=0.5 T and (b) B>6.5 T. The lattice parameter of the *c*-axis (right-hand scale) vs. temperature is also shown in (a). The shaded area highlights the region between $T_{MI}$ and $T_N$.

**Fig.2.** Isothermal magnetization M for B||a-,b-,c-axis at T=5 K. Note that the magnetic easy-axis is along the a-axis (with metamagnetic transition) with a high-field spin polarization of more than 85%.

**Fig.3.** Magnetic field dependence of the resistivity for the current along the *a*- and *c*-axis, $\rho_c$ (Fig. 3(a)) and $\rho_a$ (Fig. 3(b)), respectively, for B||*a*-, *b*- and *c*-axis at T=0.6 K.

**Fig.4.** Temperature dependence of the resistivity for the current along the *c*-axis, $\rho_c$, at a few representative B up to 28 T applied along (a) the *a*-axis and (b) the *b*-axis for 1.2 K<T<80. Inset: $\rho_c$ vs. T for $B_{||a}$=28 T. The temperature dependence of the *a*-axis resistivity $\rho_a$ is identical to that of $\rho_c$ presented here, and is therefore not shown.

**Fig.5.** M as a function of angle, θ, between the *a*-axis and B at B=6.2 T for the temperatures indicated. θ=0 corresponds to B along the *a*-axis. Note that M decreases abruptly when B rotates away from the *a*-axis for T<40 K and that the easy-axis starts to rotate to the *b*-axis (θ=90°) with increasing temperature when T>42 K.

**Fig.6.** (a) Magnetic field dependence of $\rho_c$ for B||*c*-axis for a few representative temperatures as indicated ($\rho_a$ is not shown); (b) The amplitude of the quantum oscillations (QOs) as a function of inverse field $B^{-1}$ for $\rho_c$; (c) The amplitude of the QOs as a function of inverse field $B^{-1}$ for $\rho_a$. The field dependence of the *a*-axis resistivity $\rho_a$ is nearly identical to that of $\rho_c$ presented here.

**Fig.7.** (a) Magnetic field dependence of $\rho_c$ for B (within the *ac*-plane) for various orientations, Θ, defined as the angle between B and the *c*-axis; (b) The amplitude of the quantum oscillations (QOs) as a function of inverse field $B^{-1}$ at $\theta=27^o$.

**Fig.8.** Phase diagram plotted as B vs T summarizing various phases for B||*a*-, *b*-, and *c*-axis, see panels (a), (b) and (c), respectively. Note that OO stands for orbital order, OD orbital disorder, QO for quantum oscillations, P for paramagnetism and AF for antiferromagnetism.

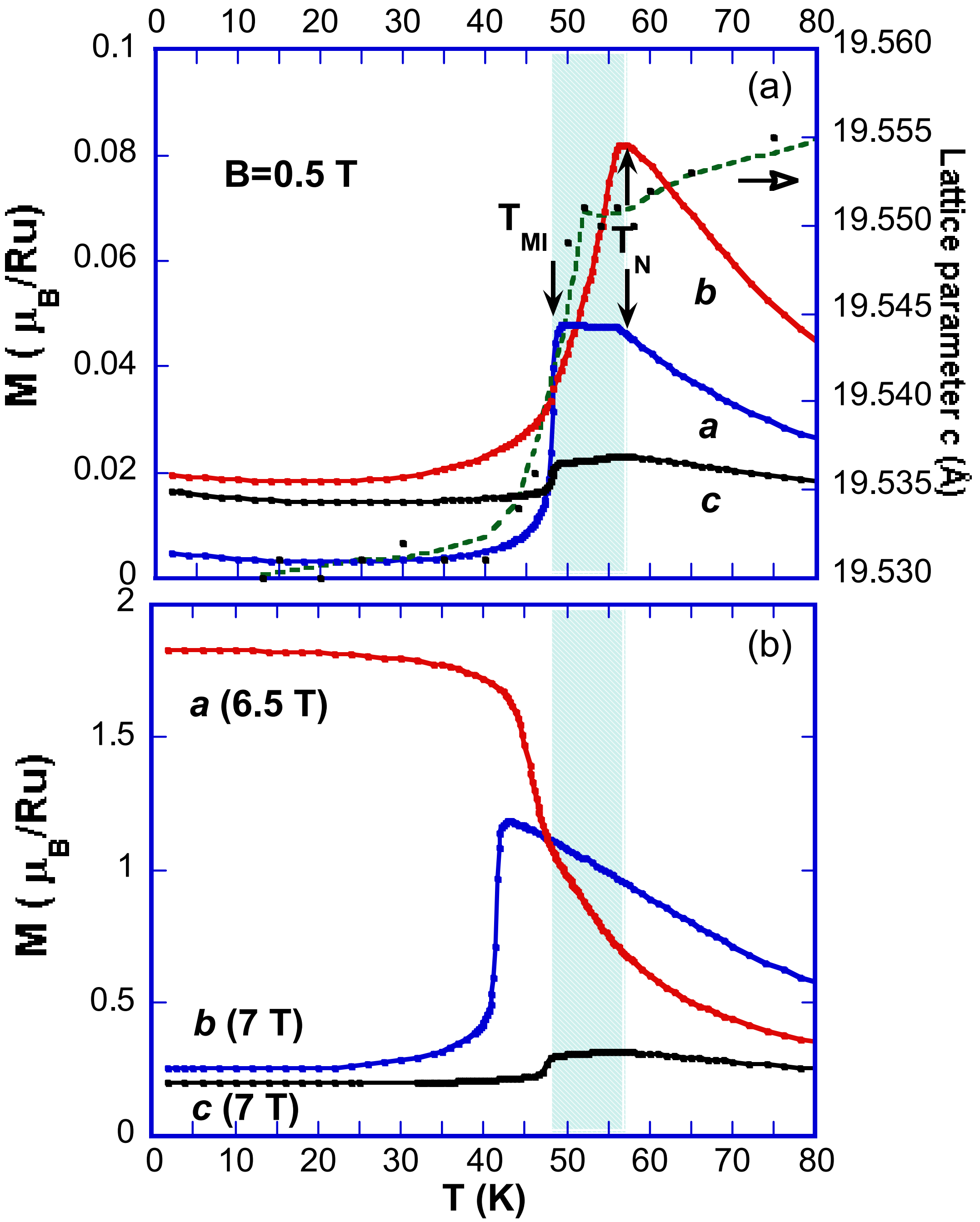

Fig.1

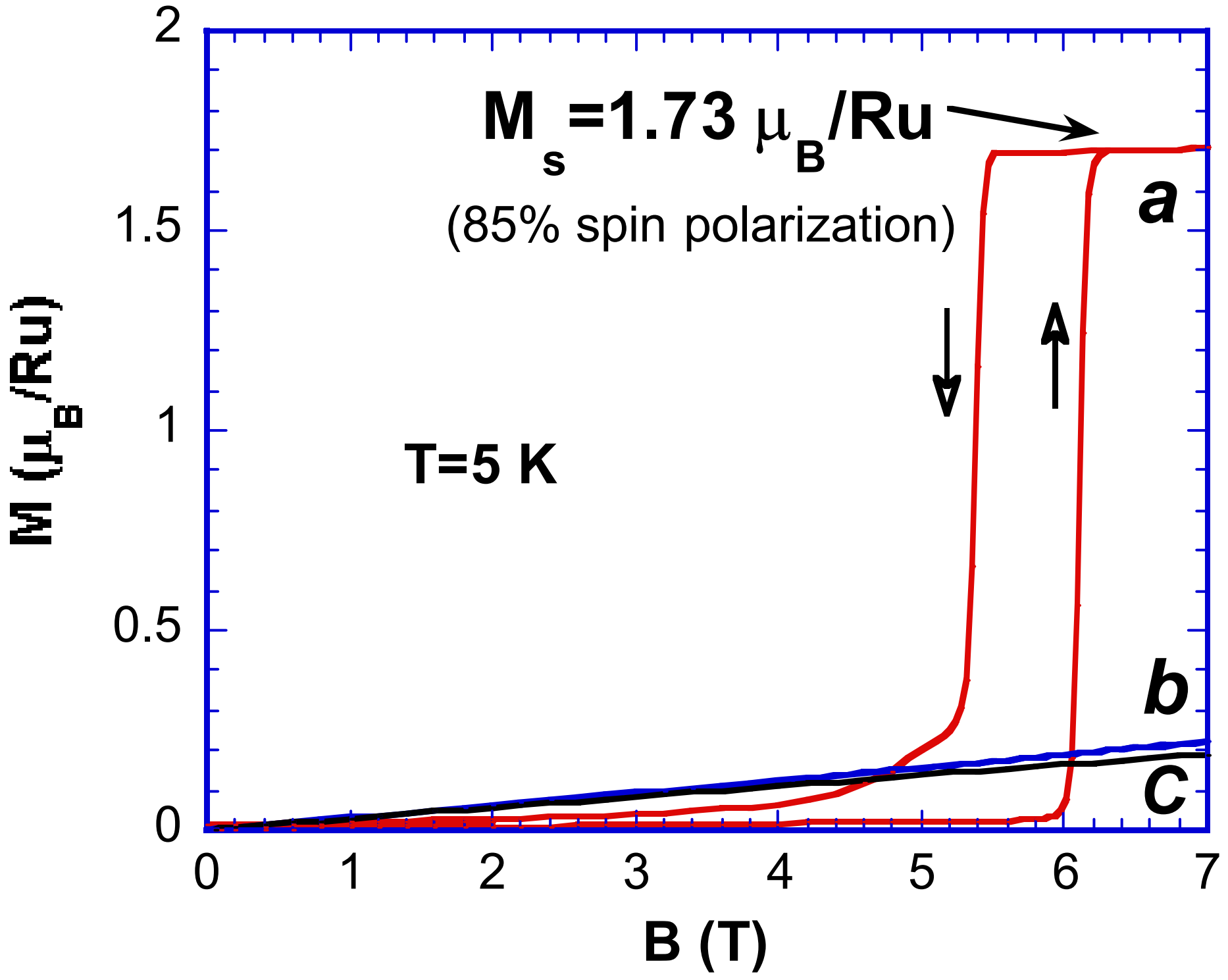

Fig. 2

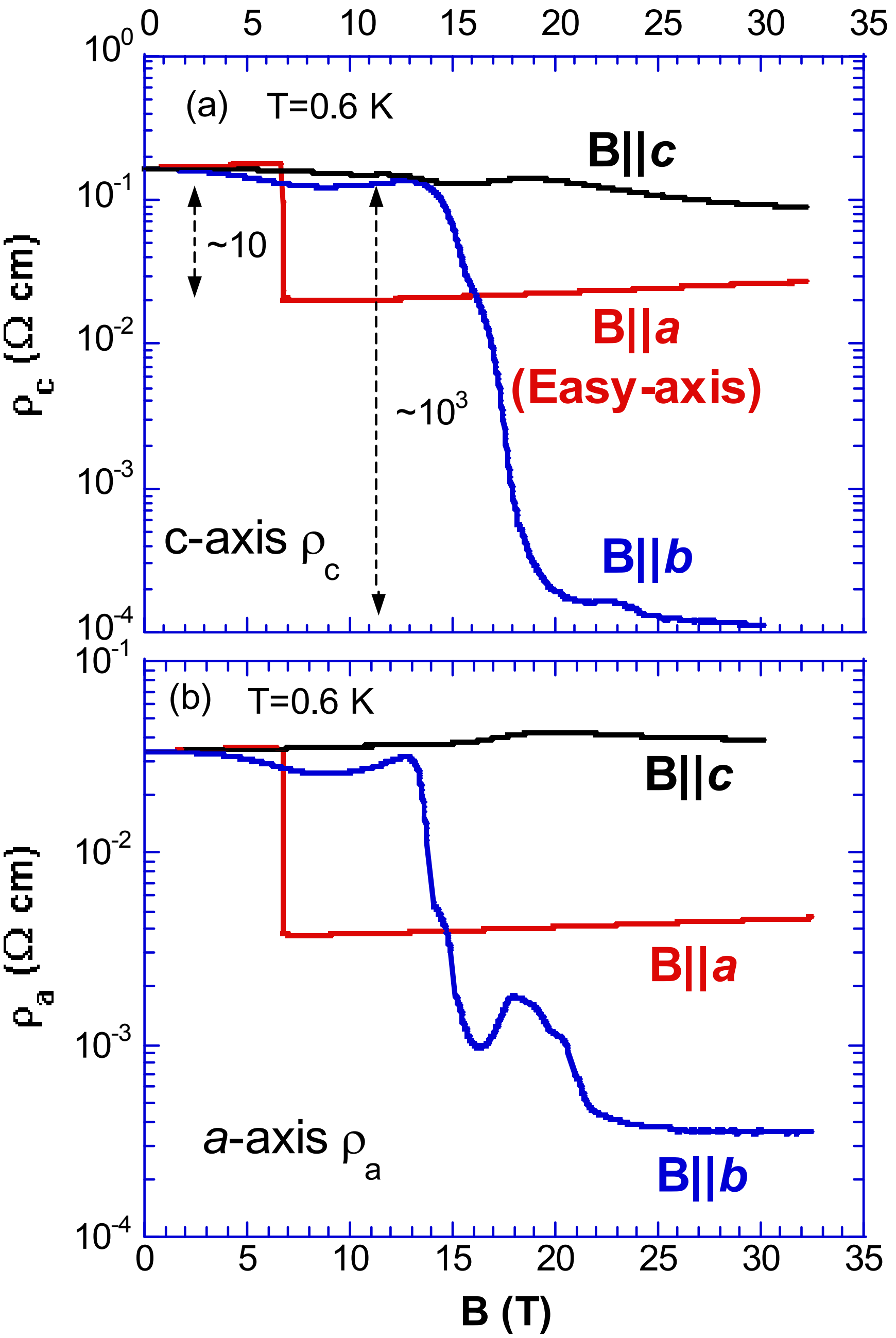

Fig. 3

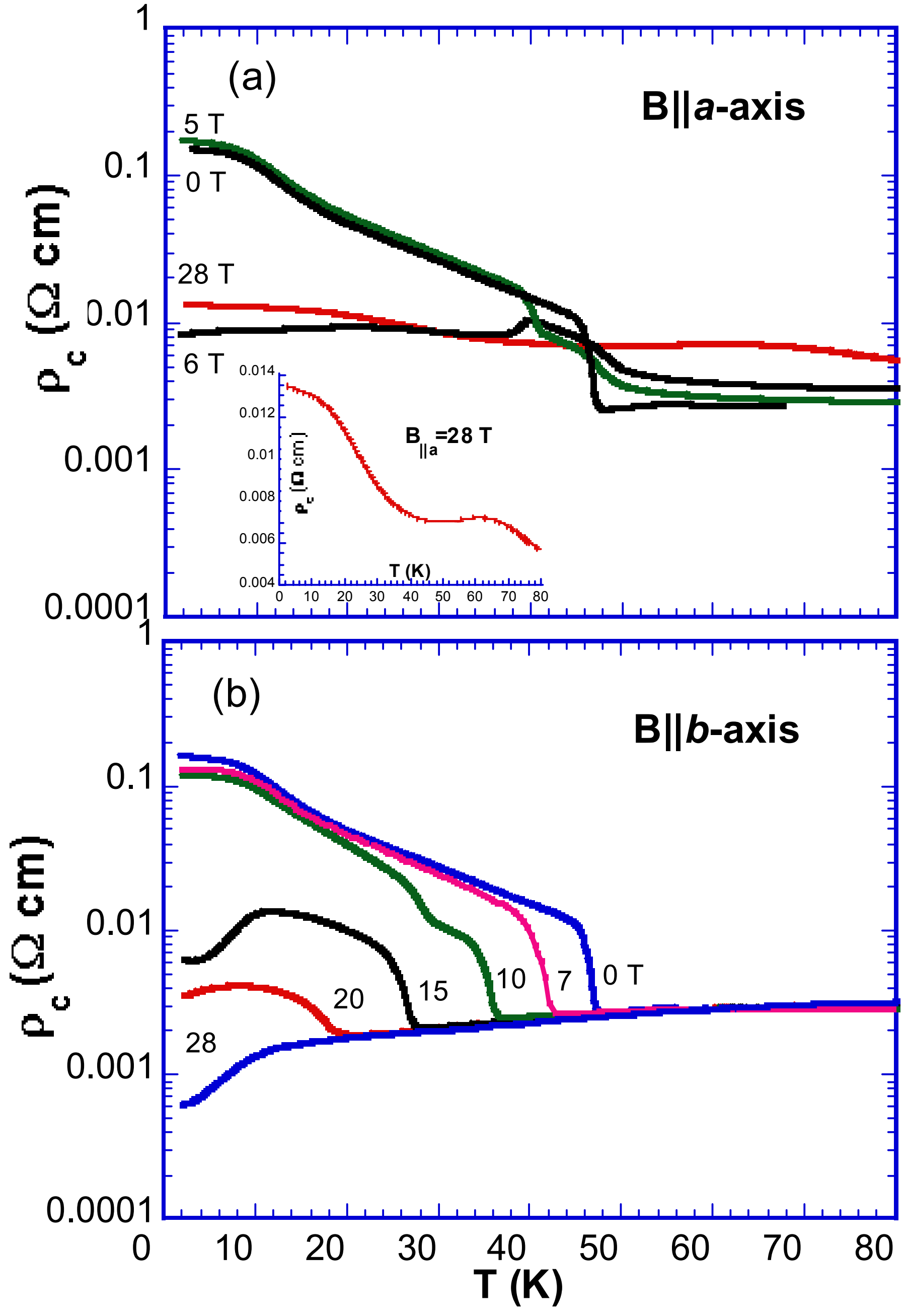

Fig. 4

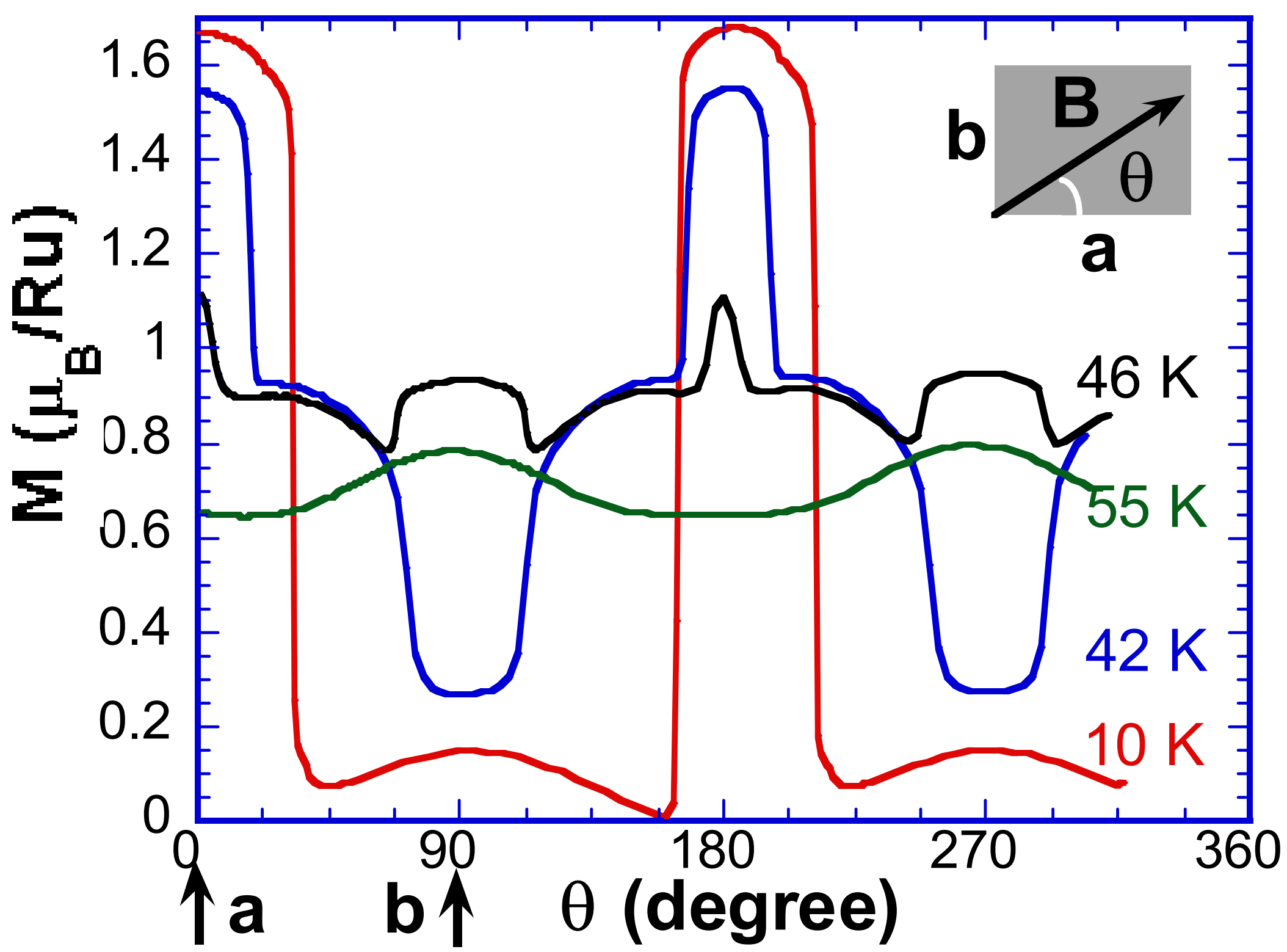

Fig. 5

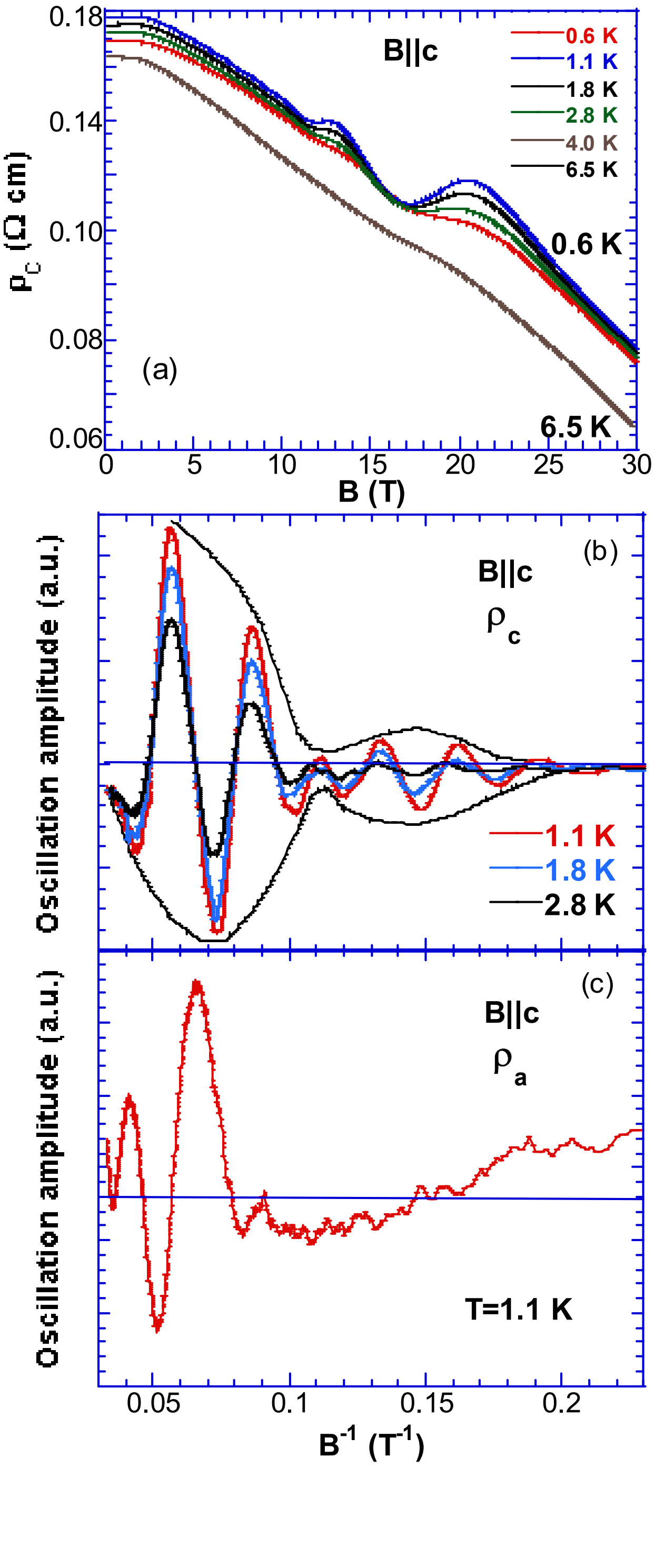

Fig.6

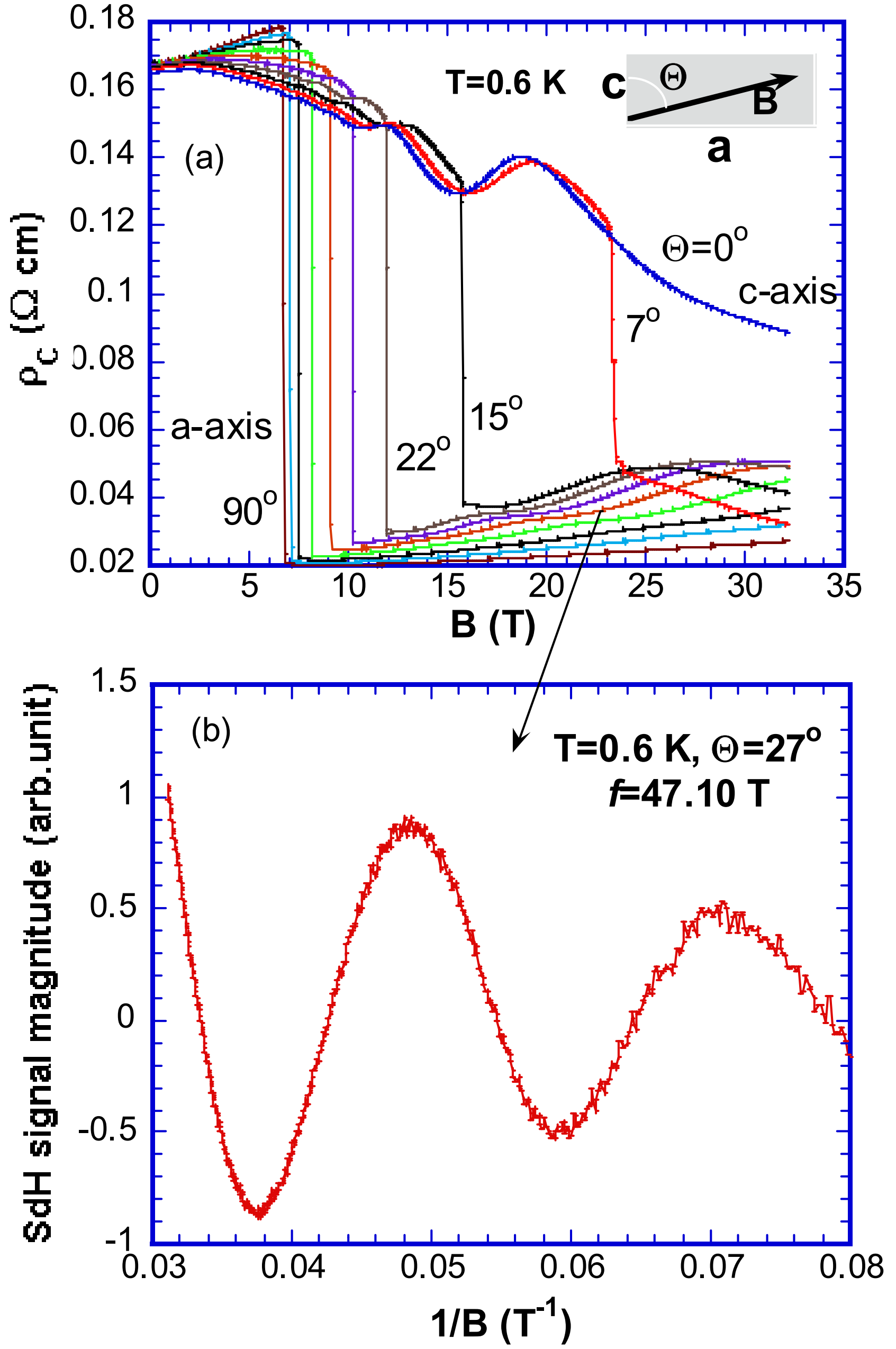

Fig.7

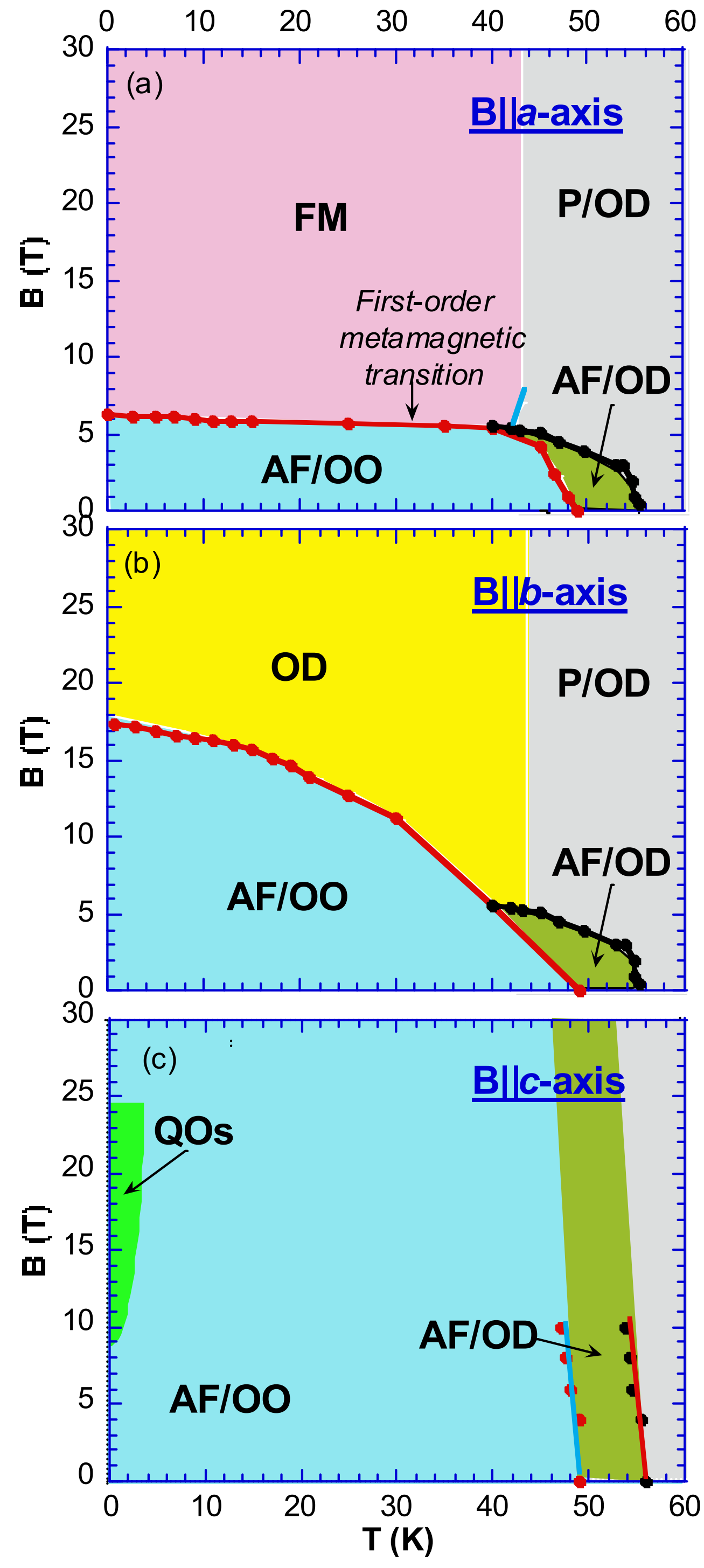

Fig.8